\documentclass{aa}
\usepackage{psfig}

\def\A{\AA}
\def\solar {\ifmmode_{\mathord\odot} \else $_{\mathord\odot}$\fi} 
\def\Msol {\ifmmode {\,{\it M}\solar} \else $\,M$\solar\fi}        
\def\simgt{\lower.5ex\hbox{$\; \buildrel\over \sim \;$}}
\def\simlt{\lower.5ex\hbox{$\; \buildrel < \over \sim \;$}}
\begin{document}


%
   \title{New Neighbours:
  II. an M9 dwarf at d~$\sim$~4~pc, DENIS-P~J104814.7$-$395606.1.
\thanks{Based on observations made at the European Southern Observatory,
and at the W.M. Keck Observatory, which is operated jointly by 
the University of California and the Californian Institute of Technology. 
Also based on plates scanned with the MAMA microdensitometer 
(http://dsmama.obspm.fr) developed and operated by
INSU/CNRS/Observatoire de Paris}}

 \author{X.~Delfosse \inst{1,3} \and T.~Forveille \inst{2,3} \and 
E.L.~Mart\'{\i}n \inst{4} \and J.~Guibert \inst{5,6} \and J.~Borsenberger
   \inst{7} \and
F.~Crifo \inst{6} \and C.~Alard \inst{6} \and N.~Epchtein \inst{8} \and 
P.~Fouqu\'e \inst{9,10} \and G. Simon \inst{6} \and F. Tajahmady \inst{5,6}
}


\institute{   Instituto de Astrof\'{\i}sica de Canarias
              E-38200 La Laguna, Tenerife, 
              Canary Islands, Spain
\and
              Canada-France-Hawaii Telescope Corporation, 
              65-1238 Mamaloha Highway,
              Kamuela, HI 96743, 
              U.S.A.
\and
              LAOG, Observatoire de Grenoble,
              BP53,
              F-38041 Grenoble,
              France
\and
              Institute for Astronomy, University of Hawaii at Manoa, 
              2680 Woodlawn Drive, 
              Honolulu, HI 96822, USA
\and
              Centre d'Analyse des Images de l'INSU, 
              61 avenue de l'Observatoire,  
              F-75014 Paris, 
              France              
\and   
              Observatoire de Paris (DASGAL/UMR-8633), 
              F-75014 Paris, France
\and 
              Institut d'Astrophysique de Paris, 
              98 bis boulevard Arago, 
              F-75014 Paris, 
              France
\and
              Observatoire de Nice, 
              BP. 4229,
              06304 Nice Cedex 4, 
              France
\and 
              DESPA, Observatoire de Paris, 
              5 place J. Janssen, 
              F-92195 Meudon Cedex, France
\and
              European Southern Observatory, 
              Casilla 19001, 
              Santiago 19, Chile
}


\offprints{Xavier Delfosse, email: delfosse@obs.ujf-grenoble.fr}

\date{Received / Accepted}


\abstract{we present the discovery of a previously unknown member of the
immediate 
solar neighbourhood, DENIS-P~J104814.7$-$395606.1 (hereafter DENIS~1048$-$39), 
identified while
mining the DENIS database for new nearby stars. A HIRES echelle 
spectrum obtained with the 10-m Keck telescope shows that it is an
M9 dwarf. DENIS~1048$-$39 has a very bright apparent magnitude 
($I=12.67$) for its spectral type and colour ($I-J=3.07$), and 
is therefore very nearby. If it is single its 
distance is only 4.1$\pm$0.6~pc, ranking it as between our 12$^{\rm th}$
and 40$^{\rm th}$ closest 
neighbour. It is also the closest star or brown dwarf with a
spectral type later than M7V. 
Its proper motion was determined through 
comparison of Sky atlas Schmidt plates, scanned by the MAMA microdensitometer,
with the DENIS images. At 1.52$"\,$yr$^{-1}$ it primarily attests the
closeness of DENIS~1048$-$39 and hence its
dwarf status.
These characteristics make it an obvious
target for further detailed studies.
}

\maketitle

\titlerunning{DENIS-P~J104814.7$-$395606.1: An M9 dwarfs at d~$\sim$~4~pc}
\authorrunning{Delfosse et al.}

%

\section{Introduction}

Much of our understanding of stellar astronomy rests upon the nearest 
stars. As individual objects they are the brightest and hence best
studied examples of their spectral type, and they have distances that 
can usually be measured directly from an accurate trigonometric parallax.
These stars are also the source of some of the most accurate stellar 
masses: for nearer multiple 
systems the same physical separation 
translates into a wider angular separation, and hence into a better
characterized orbit. As a population, the solar neighbourhood sample
therefore provides deep insight into the nature of our Galaxy's components,
through studies of its stellar luminosities and mass functions,
its kinematics, its chemical composition, and its multiplicity statistics.

For the moderately bright ({M$_V$}$<$7) G and K dwarfs, our current 
census of the solar neigbourhood is complete out to at least 25~pc 
(Jahreiss \cite{jahr94}). Intrinsically fainter stars however dominate the 
galactic population budget, and represent a large fraction of its
mass budget: M dwarfs account for at least 70\% of all stars, and
about 50\% of the mass of the galactic disk. For this fainter population
our present census unfortunately becomes incomplete at a rather small 
distance: from a comparison of observed star densities within 5 and 10~pc,
Henry et al. (\cite{henry97}) demonstrate that over 100 systems are missing
in the larger volume, even under the purposedly optimistic assumption
that the current inventory is complete for the 5~pc sphere. In addition 
to these systems which are altogether missing, a further source of
incompleteness is that some of the identified nearby stellar systems have 
unrecognized components, as illustrated by the sustained discovery
rate of new companions to nearby M dwarfs (e.g. Henry et al. \cite{henry99}; 
Delfosse et al. \cite{delfosse99a}; Beuzit et al. 
in preparation).

\begin{table}
\tabcolsep 1.1mm
\begin{tabular}{lllll} \hline
\multicolumn{5}{c}{Photometry} \\
B=19.0 &  R=15.7 & I=12.67 & J=9.59 & K$_s$=8.58 \\ 
$\pm$0.3 & $\pm$0.2 & $\pm$0.05 & $\pm$0.05 &  $\pm$0.05   \\ \hline
\multicolumn{5}{c}{Astrometry} \\
$\alpha$ (2000.0) & $\delta$ (2000.0) & $\mu$ $\alpha$ ($"\,$yr$^{-1}$) & $\mu$ $\delta$
($"\,$yr$^{-1}$) & \\ 
10:48:14.68 &  $-$39:56:06.1 & $-$1.147$\pm$0.008 & $-$0.992$\pm$0.009 & \\ \hline
\end{tabular}
\caption{Basic parameters of DENIS~1048$-$39. The 
photometry is extracted from the DENIS catalog for the I, J and K$_s$
bands and from the sky atlas Schmidt plates for the B and R bands. The 
J2000.0 position of the object is for epoch 1999.11 (date of the DENIS
observations).}
\label{table_para}
\end{table}

To address both of these incompleteness sources, some of us are
conducting a systematic search for companions to 
nearby M dwarfs (Delfosse et al. \cite{delfosse99a},
for a presentation of that programme), and we use the DENIS near infrared sky 
survey to complete the identification of the southern solar neighbourhood
M dwarfs.
Here we present an initial result of that second programme, the discovery 
and initial characterization of 
a previously unknown very nearby (d$\sim$~4~pc) M9 dwarf.

\section{Observational data}


The DEep Near-Infrared Survey (DENIS) is a southern sky 
survey (Epchtein \cite{epchtein97}), which will provide
full coverage of the southern hemisphere in two near-infrared bands (J
and K$_s$) and one optical band (I).
DENIS observations are carried out on the ESO 1m telescope at La Silla.
Dichroic beam splitters separate the three channels, and
focal reducing optics provide scales of 3$''$ per pixel on the
256$\times$256 NICMOS3 arrays used for the two infrared channels and
1$''$ per pixel on the 1024$\times$1024 Tektronix CCD detector of the I channel.
The sky is scanned in a step and stare mode, along
30 degrees strips at constant right ascension.

The image data were processed with the
standard pipeline software (Borsenberger \cite{borsen97}, 
Borsenberger et al., in preparation) at the 
Paris Data Analysis Center (PDAC).  The instrumental and
sky background are derived from a local clipped mean along the
strip. Flat-field corrections are derived from observation of the
sunrise sky.
Source extraction and photometry is performed at PDAC, using 
a space-varying kernel (Alard \cite{alard00}) algorithm. The astrometry
of the individual DENIS frames is referenced to the USNO-A2.0 catalog,
whose $\sim$~1$''$ accuracy therefore determines the absolute
precision of the DENIS positions.

With typical 3-$\sigma$ limits of I$\sim$19.0, J$\sim$16.0 and 
K$_s\sim$13.5, the DENIS survey is very sensitive to very
late M and L dwarfs (Delfosse et al. \cite{delfosse99b}; Mart\'{\i}n 
et al. \cite{martin99}). We are extracting large statistical sample 
of such very low mass stars and brown dwarfs, by selecting sources
with $I-J \geq 3.00$ (later than M8) and $|l_{II}| \geq 20^{\circ}$ 
(Delfosse et al., in preparation). A set of morphological parameters 
is used to reject artefacts. 

While analysing the first 3000 square degrees processed by PDAC, we 
have identified a bright and previously unknown red object ($I-J=3.07$), 
DENIS-P~J104814.7$-$395606.1 (hereafter DENIS 1048$-$39). Its DENIS 
photometry (Table~\ref{table_para}) 
is most consistent with a very close M8-M9 dwarf, and somewhat less
compatible with a cool giant at halo-like distances.


\begin{figure}
\psfig{height=6.3cm,file=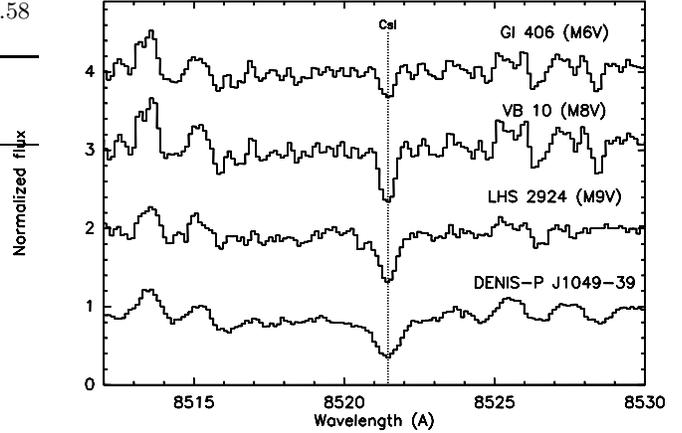,angle=-90} 
\caption{8521~\A~CsI line of very late 
M dwarf templates and DENIS 1048$-$39.}
\label{fig_spect}
\end{figure}

DENIS 1048$-$39 was observed with the HIRES 
echelle spectrograph (Vogt et al. \cite{vogt94}) 
on the 10~m Keck~I telescope (Mauna 
Kea, Hawaii), on the nights of May 30$^{\mathrm{th}}$ and 31$^{\mathrm{st}}$
2000, to obtain a 
high-resolution spectrum (R=31,000). Gl~406, VB10 and LHS2924 were
observed with the same instrumental configuration, for use as late M-type 
spectral templates.

Figure~\ref{fig_spect} shows the spectrum of DENIS 1048$-$39 
(and the spectral templates) around the 8521~\A~CsI resonance doublet. 
The detection of a strong CsI line immediately demonstrates that
DENIS 1048$-$39 is a dwarf, as this feature disappears
at the lower gravity of a giant's atmosphere. Basri et al. 
(\cite{basri00}) have demonstrated that alkali resonance doublets 
are excellent effective temperature diagnosis. We therefore use
the equivalent width of the CsI line (Table~\ref{table_spect}) to
determine a spectral type of M9V for DENIS 1048$-$39,
through comparison with the spectral templates. This determination
coincides with the result of an eyeball comparison of the spectra,
and we estimate that it is accurate to better than 0.5 subclass.

We note that H$_\alpha$ emission 
is variable. On the first night we did not detect H$_\alpha$.  
We set an upper limit to the equivalent width of 0.11 \A. However,  
on the second night we detected H$_\alpha$ emission with equivalent width 
of 1.1 \A. This is a low level of H$_\alpha$ emission for an M9 dwarf
(Gizis et al. \cite{gizis00}). 
We do not detect the lithium resonance feature at 670.8~nm, with an 
upper limit to its equivalent width of 0.12 \A. This implies substantial 
lithium depletion with respect 
to the nearby M9 brown dwarf LP 944$-$20, for which Tinney 
(\cite{tinney98}) measured 
EW(LiI)=0.53 \A. Thus, DENIS 1048$-$39 is more massive and older 
than LP 944$-$20. It will be interesting to compare gravity sensitive 
spectral features in the two objects. 
We have also measured the heliocentric radial velocity and 
rotational broadening of DENIS 1048$-$39. The results are 
$V_R$=-10.1$\pm$0.5 km\,s$^{-1}$ 
and $v\,\sin (i)$=27$\pm$5 km\,s$^{-1}$. Both measurements were derived by 
cross-correlation 
with the spectrum of VB~10 obtained on the same night and with the same 
instrumental configuration. We adopted a radial velocity of 
34.4 km\,s$^{-1}$ for VB~10 (Tinney \& Reid \cite{tinney98b}).

\begin{table}
\begin{tabular}{lll} \hline
Object & Spectral Type & EW(CsI) \\ \hline \hline
Gl~406 & dM6 & 0.21$\pm$0.02 \\
VB10   & dM8 & 0.38$\pm$0.03 \\
LHS2924  & dM9 & 0.42$\pm$0.04 \\
DENIS 1048$-$39 & dM9$^{\rm a}$ & 0.65$\pm$0.06 \\
LP944$-$20 & bdM9 & 0.60$^{\rm b}$ \\ \hline
\end{tabular}
\caption{Equivalent width of CsI line (8521~\A) for the very late 
M dwarf templates and for DENIS 1048$-$39. (a) The spectral 
type of the DENIS object is determined from the present observations.
(b) Value from Basri et al. \cite{basri00}}
\label{table_spect}
\end{table}


\begin{figure}
\psfig{height=4.0cm,file=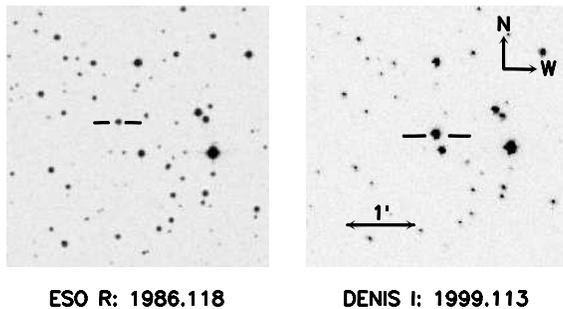,angle=-90} 
\caption{The position of DENIS 1048$-$39 in the ESO-R 
and DENIS-I image.}
\label{fig_carte}
\end{figure}

A very bright M9 dwarf such as DENIS 1048$-$39
has to be very close, and it must therefore have a large proper
motion, unless its tangential velocity serendipitously
happens to be very small.

The DENIS object is easily seen on archival plates (field 318) from the
POSS I (1964.209), the SRC-J 
atlas (epoch 1975.364), the ESO-R atlas (epoch 1986.118) and the 
SRC-R atlas (epoch: 1993.463). Figure~\ref{fig_carte} shows the
motion of DENIS 1048$-$39 over the last 13 years.
We used the MAMA microdensitometer (Berger et al. \cite{berger91}) to
digitize the ESO-R and SRC-J survey plates. The astrometry and
the photometry were respectively referenced to the ACT
(Astrographic Catalogue $+$ Tycho; Urban et al. \cite{urban98}) 
and to the GSPC-2 (Postman et al. 
\cite{postman97}). The position on the SRC-R survey plate was taken from 
the APM catalog, and for the POSS I plates it was measured with 
SeXtractor from an image obtained from the DSS scan server, 
and referenced to several surrounding stars. From the five available 
epochs (including the DENIS position at epoch 1999.113), we determine a 
total proper motion of 1.52$"\,$yr$^{-1}$ (Table \ref{table_para}) with an
accuracy of 0.009$"$.yr$^{-1}$ on each axis. 
Such a proper motion would easily have qualified DENIS 1048$-$39 for
inclusion in the LHS catalogue (Luyten \cite{luyten79}). 
The absence of this relatively
bright star (B=19.0, R=15.7, I=12.7) again illustrates
the acknowledged incompleteness of that catalog below -33$^\circ$,
the declination limit of the Palomar Sky Surveys. It also confirms 
the dwarf status of DENIS 1048$-$39.

\section{Discussion}

Only five M9 dwarfs have published trigonometric parallaxes
(Kirkpatrick et al. \cite{kir00} for a review). Of those, LP944$-$20 is 
a young brown dwarfs and its absolute magnitudes and colours
are slightly different from the others. Using the four stellar M9 dwarfs,
we determine mean absolute magnitudes of $M_I=14.77\pm0.2$, 
$M_J=11.53\pm0.2$ and $M_K=10.30\pm0.2$. If DENIS 1048$-$39 is a single 
star, its I, J and K magnitudes therefore translate into distances
of respectively $d_I=3.8\pm0.35$pc, $d_J=4.1\pm0.35$pc and
$d_K=4.5\pm0.35$pc. The slight inconsistency between the three distances
may be due in part to our comparing the photometry of DENIS 1048$-$39 in the 
DENIS system with absolute magnitudes referenced to the Cousins-CIT 
system. A preliminary comparison of the two photometric systems 
for very late M dwarfs (Delfosse \cite{delfosse97}) shows 
$\sim$0.1~mag differences in the J and K bands. Any difference in the I band
on the other hand is below 0.05~mag, and the distance determined from the 
I magnitude should thus be the most precise. For the time being, we 
nonetheless prefer to consider the issue as incompletely resolved
and we conservatively adopt $d=4.1~\pm~0.6$~pc. 
If DENIS 1048$-$39 is actually an unresolved double star
it is slightly more distant, with the maximum distance of 
$d~\sim~5.8$~pc reached for an equal mass binary. 

The single-star distance would make DENIS~1048$-$39 our 12$^{\rm th}$ to 
37$^{\rm th}$ nearest neighbour system, while the larger distance for
an equal mass binary would place it around the 60$^{\rm th}$ place. 
If single it is also the closest dwarf later than M7V (Table~\ref{tab_neigh}),
and whatever its multiplicity status it is the brightest very late M 
dwarf. As such DENIS 1048$-$39 
is clearly destined to become an observational benchmark. 
At $I=12.7$ it is in particular bright enough for very high signal to noise 
high resolution spectroscopy. 

The $UVW$ heliocentric space motions, calculated from the above proper 
motion, radial 
velocity, and distance are $U=11$, $V=0$ and $W=-29$ km\,s$^{-1}$,
(with $U$ defined as positive away from the Galactic Center). 
They are typical of the young disk dynamical population.

It is clearly essential to measure the trigonometric parallax of this 
new nearby star. This can be achieved with useful precision even
within a single observing season, since this parallax will range 
between 300 and 200 milliarcseconds, depending on the actual
multiplicity of the system. Whether it is multiple however may be resolved
even before its parallax
can be determined, since it is easily bright enough (I=12.7) for
adaptive optics imaging.

\begin{table}
\begin{tabular}{llll} \hline
Objects & Spect. & Dist.& Method\\ 
        & type     & (in pc)            & \\ \hline \hline
DENIS 1048$-$39 & M9 & 4.1 & spect. par.\\
LP944$-$20 & M9$^{\rm a}$ & 4.96$^{\rm b}$ & trigo. par.\\
DENIS-P J0255$-$4700 & L6$^{\rm c}$ & 5.0$^{\rm d}$ & spect. par.\\
Gl 644D (VB8) & M7$^{\rm e}$ & 5.74$^{\rm f}$  & trigo. par.\\
Gl 229B & T$^{\rm g}$ & 5.77$^{\rm f}$ & trigo. par.\\
Gl 752B (VB10) & M8$^{\rm e}$ & 5.83$^{\rm h}$  & trigo. par.\\
Gl 570D & T$^{\rm i}$ & 5.91$^{\rm f}$ & trigo. par.\\
2MASSI J0559191$-$140448 & T$^{\rm j}$ & 6.0$^{\rm d}$ & spect. par.\\
GJ 3877 (LHS 3003) & M7$^{\rm e}$ & 6.36$^{\rm k}$ & trigo. par.\\ \hline
\end{tabular}
\caption{M, L and T dwarfs of spectral type later than 
M7 and closer than 7~pc. Note that spectral type of
DENIS-P J0255$-$4700 is in the Mart\'{\i}n et al. (\cite{martin99}) 
system. The references are:
(a) Kirkaptrick et al. (\cite{kir99}); (b) Tinney (\cite{tinney96});
(c) Mart\'{\i}n et al. (\cite{martin99}); (d) Kirkpatrick et al. 
(\cite{kir00});
(e) Kirkpatrick et al. (\cite{kir95}); (f) ESA (\cite{esa97}) 
(Hipparcos catalog); 
(g) Nakajima et al., (\cite{naka95}); (h) Van Altena et al. 
(\cite{vanaltena95});
(i) Burgasser et al., (\cite{burg00a}); (j) Burgasser et al., 
(\cite{burg00b}); (k) Ianna (\cite{ianna93})}
\label{tab_neigh}
\end{table}

The discovery of this very close M9 dwarf shows again, if need there be,
that our inventory of the immediate solar neighbourhood is 
incomplete and misses isolated very low mass objects, in addition to
unrecognised members of identified systems. Table~\ref{tab_neigh} lists 
the 9 objects
with a spectral type later than M7 that are known within 7~pc. 
One can note that 4 of these have been discovered during the past years, 
by either the 2MASS (Gl~570D and 2MASSI~J0559191$-$140448) or the DENIS 
(DENIS 1048$-$39 and DENIS-P~J0255$-$4700) surveys. The approaching
completion of these surveys, and their subsequent full analysis, will 
bring us much closer to a full census of our neighbourhood. 

\begin{figure}
\psfig{height=5.0cm,file=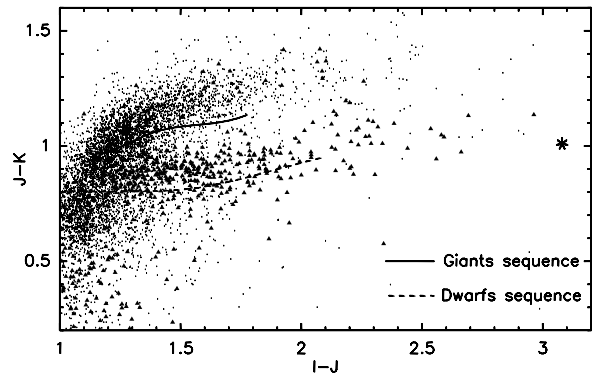,angle=-90}
\caption{All DENIS sources detected in the first 3000 squares degrees 
processed by PDAC with a photometric parallax $\leq$ 25~pc, galactic
latitude ${\geq}~30^\circ$ and $I-J$ in the 1.0-3.0 range (corresponding
to the M spectral class). 
The photometric parallaxes were obtained from an ${\mathrm M_I(I-J)}$
relation apropriate for dwarfs. The dots represent the object which have a
counterpart in the USNO A2 catalog, while the triangles have no counterpart
(and include all high proper motion stars). The asterisk represents
DENIS-P J1048$-$39. The solid and dashed lines respectively
represent the giant and dwarf sequences of Bessel \& Brett (1988),
for slightly different filters.
}
\label{ij_jk}
\end{figure}

Specifically,
we are now deriving photometric parallaxes from the I and J magnitudes of
all DENIS sources at high galactic latitude. From this we assemble a 
sample of
candidate members of the solar neighbourhood. That sample still contains
some contamination by giants, which from the DENIS I$-$J/J$-$K colour-colour
diagram alone can be reduced significantly but not excised with 100\% 
completeness and reliability (Fig.~\ref{ij_jk}). As an initial step towards 
rejecting
giants, we determine the proper motion of the candidates from a comparison
of the DENIS position with archival Schmidt plates digitized on the MAMA 
microdensitometer. Our eventual goal however is to obtain spectra of
all candidates, independently of their proper motions, to produce
a complete inventory of the solar neighbourhood for the high galactic
latitude southern sky that has no kinematic bias. 
In a second stage, we plan to extend the search to include the low galactic 
latitudes, where contamination by giant or dust-reddened stars is much more
severe and precludes obtaining spectra all photometric candidates. 
We will there resort to proper motion selection amongst these candidates, 
and will use the high galactic latitude sample to measure the resulting
kinematic bias.

%


\begin{acknowledgements}

We thank the CAI-MAMA team for digitizing the ESO and SRC Schmidt plates. 
We are very grateful to R.D. Scholz for helping us find additional old 
plates that detect DENIS 1048$-$39, and to the referee, C. Leinert, for 
several suggestions that improved the paper. 
The DENIS project is supported by the {\it SCIENCE  and the Human Capital and
Mobility} plans of the European Commission under grants CT920791 and CT940627,
by the French INSU, the Minist\`ere de
l'Education Nationale and the CNRS, in
Germany by the State of Baden-W\"urtemberg, in Spain by the DGICYT, 
in Italy by
the CNR, by the Austrian Fonds zur F\"orderung
der wissenschaftlichen Forschung und Bundesministerium f\"ur Wissenschaft und
Forschung and the ESO C \& EE grant A-04-046.

\end{acknowledgements}


\begin{thebibliography}{}

\bibitem[2000]{alard00} Alard C., 2000, A\&AS 144, 363

\bibitem[2000]{basri00} Basri G., Mohanty S., Allard F., et al.,
2000, ApJ 538, 363

\bibitem[1988]{bessell88} Bessell M.S., Brett J.M., 1988,
PASP 100, 1134


\bibitem[1991]{berger91} Berger J., Cordoni J.-P., Fringant A.-M., et al.,
1991, A\&AS 87, 389 

\bibitem[1997]{borsen97} Borsenberger J., 1997, in
The impact of large scale near-infrared surveys, 
F. Garzon et al. (eds)
Kluwer Dordrecht

\bibitem[2000a]{burg00a} Burgasser A.J., Kirkpatrick J.D., Cutri R.M., 
et al., 2000a, ApJ 531, L57

\bibitem[2000b]{burg00b} Burgasser A.J., Wilson J.C., Kirkpatrick J.D.,
et al., 2000b, AJ 120, 1000

\bibitem[1997]{delfosse97} Delfosse X., 1997, PhD thesis, Grenoble University

\bibitem[1999a]{delfosse99a} Delfosse X., Forveille T., Beuzit J.-L.,
et al., 1999a, A\&A 344, 897

\bibitem[1999b]{delfosse99b} Delfosse X., Tinney C.G., Forveille T.,
et al., 1999b, A\&AS 135, 41


\bibitem[1997]{epchtein97} Epchtein N., 1997, in 
The impact of large scale near-infrared surveys, 
F. Garzon et al. (eds)
Kluwer Dordrecht

\bibitem[1997]{esa97} ESA, 1997, The HIPPARCOS Catalogue, ESA SP-1200

\bibitem[2000]{gizis00} Gizis J.E., Monet D.G., Reid I.N., et al., 
2000., AJ 120, 1085

\bibitem[1997]{henry97} Henry T.J., Ianna P.A., Kirkpatrick J.D., 
Jahreiss H., 1997, AJ 114, 388

\bibitem[1999]{henry99} Henry T.J., Franz O.G., Wasserman L.H., et al.,
1999, ApJ 512, 864

\bibitem[1993]{ianna93} Ianna P.A., 1993, in Developments in Astrometry
and Theirs Impact on Astrophysics and Geodynamics, I.I. Mueller and
B.Kolaczek (eds), Kluwer Boston, 75

\bibitem[1994]{jahr94} Jahreiss H, 1994, Ap{\&}SS, 217, 63

\bibitem[1995]{kir95} Kirkpatrick J.D., Henry T.J., Simons D.A., 1995,
AJ 109, 797

\bibitem[1999]{kir99} Kirkpatrick J.D., Reid I.N., Liebert J., et al.,
1999, ApJ 519, 802

\bibitem[2000]{kir00} Kirkpatrick J.D., Reid I.N., Liebert J., et al., 
2000, AJ 120, 447

\bibitem[1979]{luyten79} Luyten W.J., 1979, LHS catalogue, Minneapolis,
University of Minnesota, 2$^{\rm nd}$ edition
 
\bibitem[1999]{martin99} Mart\'{\i}n E.L., Delfosse X., Basri G., et al.,
1999, AJ 118, 2466

\bibitem[1992]{monet92} Monet D.G., Dahn C.D., Vrba F.J., et al., 1992,
AJ 103, 638

\bibitem[1995]{naka95} Nakajima T., Oppenheimer B.R., Kulkarni S.R.,
et al., 1995, Nature 378, 463

\bibitem[1997]{postman97} Postman M., Bucciarelli B., Sturch C. et al.,
1997, IAU Symposium 179, 379 


\bibitem[1996]{tinney96} Tinney C.G., 1996, MNRAS 281, 644

\bibitem[1998]{tinney98} Tinney C.G., 1998, MNRAS 296, L82

\bibitem[1998]{tinney98b} Tinney C.G, Reid I.N., 1998, 
MNRAS 301, 1031

\bibitem[1998]{urban98} Urban S.E., Corbin T.E., Wycoff G.L, 1998, AJ 115,
2161            

\bibitem[1995]{vanaltena95}  Van Altena W.F., 
Lee J.T., Hoffleit D., 1995, The General Catalogue of Trigonometric Stellar 
Parallaxes, Fourth edition, Yale University Observatory

\bibitem[1994]{vogt94} Vogt S.S., Allen S.L., Bigelow B.C., 1994,
in Instrumentation in Astronomy III, Proc. SPIE, 2198, 362, 
ed. Crawford, D.L., Craine, E.R.

\end{thebibliography}
\end{document}